\documentclass[conference]{IEEEtran}
\IEEEoverridecommandlockouts
\usepackage{cite}
\usepackage{amsmath,amssymb,amsfonts}
\usepackage{algorithmic}
\usepackage{graphicx}
\usepackage{textcomp}
\usepackage{xcolor}
\usepackage[inline]{enumitem}
\usepackage{subcaption}
\usepackage{easyReview}
\usepackage{array}
\usepackage{float}

\def\BibTeX{{\rm B\kern-.05em{\sc i\kern-.025em b}\kern-.08em
    T\kern-.1667em\lower.7ex\hbox{E}\kern-.125emX}}

\usepackage[acronym,nopostdot,  style=super, nonumberlist, toc]{glossaries}
\newacronym[plural=HITLs,firstplural=human-in-the-loop (HITL)]{HITL}{HITL}{human-in-the-loop}
\newacronym[plural=CPSs,firstplural=cyber-physical systems (CPS)]{CPS}{CPS}{cyber-physical system}
\newacronym[]{TSN}{TSN}{time sensitive networking}
\newacronym[plural=CPSs]{UE}{UE}{user equipment}
\newacronym{HARQ}{HARQ}{Hybrid-ARQ}
\newacronym{MCS}{MCS}{modulation coding scheme}
\newacronym{MLP}{MLP}{multi-layer perceptron}
\newacronym{SDR}{SDR}{software-defined radio}
\newacronym{OAI}{OAI}{OpenAirInterface}
\newacronym{SINR}{SINR}{signal to interference and noise ratio}
\newacronym{CDF}{CDF}{cumulative distribution function}
\newacronym{RTT}{RTT}{round-trip time}
\newacronym{PTP}{PTP}{precision time protocol}
\newacronym{TDD}{TDD}{time division duplex}
\newacronym[plural=RANs,firstplural=radio access networks (RAN)]{RAN}{RAN}{radio access network}
\newacronym{COTS}{COTS}{commercial off-the-shelf}
\newacronym[plural=PDFs,firstplural=probability density functions (PDF)]{PDF}{PDF}{probability density function}

\usepackage{amsmath,amsfonts} 
\usepackage{graphicx}

\begin{document}





\title{ExPECA: An Experimental Platform for Trustworthy Edge Computing Applications}

\author{
    \IEEEauthorblockN{Samie Mostafavi\IEEEauthorrefmark{1}, Vishnu Narayanan Moothedath\IEEEauthorrefmark{2}, Stefan Rönngren\IEEEauthorrefmark{3}, Neelabhro Roy\IEEEauthorrefmark{4},\\ Gourav Prateek Sharma\IEEEauthorrefmark{5}, Sangwon Seo\IEEEauthorrefmark{6}, Manuel {Olguín Muñoz}\IEEEauthorrefmark{7}, and James Gross\IEEEauthorrefmark{8}}
    \IEEEauthorblockA{\text{KTH Royal Institute of Technology}, Stockholm, Sweden \\
    \{\IEEEauthorrefmark{1}ssmos, \IEEEauthorrefmark{2}vnmo, \IEEEauthorrefmark{3}steron, \IEEEauthorrefmark{4}nroy,  \IEEEauthorrefmark{5}gpsharma, \IEEEauthorrefmark{6}sangwon, \IEEEauthorrefmark{8}jamesgr\}@kth.se,
    \IEEEauthorrefmark{7}manuel@olguinmunoz.xyz \\}
     
    https://expeca.proj.kth.se
}

\maketitle

\thispagestyle{plain}
\pagestyle{plain}

\begin{abstract}
This paper presents ExPECA, an edge computing and wireless communication research testbed designed to tackle two pressing challenges: comprehensive end-to-end experimentation, and high levels of experimental reproducibility.
Leveraging Openstack-based Chameleon Infrastructure (CHI) framework for its proven flexibility and ease of operation, ExPECA is located in a unique, isolated underground facility, providing a highly controlled setting for wireless experiments.
The testbed is engineered to facilitate integrated studies of both communication and computation, offering a diverse array of Software-Defined Radios (SDR) and Commercial Off-The-Shelf (COTS) wireless and wired links, as well as containerized computational environments.
We exemplify the experimental possibilities of the testbed using OpenRTiST, a latency-sensitive, bandwidth-intensive application, and analyze its performance.
Lastly, we highlight an array of research domains and experimental setups that stand to gain from ExPECA's features, including closed-loop applications and time-sensitive networking.

\end{abstract}


\section{Introduction}

As we move onto the next decade, edge computing applications are poised for significant growth, fueled by advancements in wireless technologies such as 5G and Beyond 5G (B5G), as well as breakthroughs in compute-intensive tasks like model predictive control, multimedia processing, and machine learning algorithms.
In order to achieve trustworthiness and dependability in these interdisciplinary, distributed systems, involving computer science and wireless research, it is imperative to develop sophisticated scientific tools.
On one hand, research in areas like wireless networking methods necessitates close interaction with radio hardware and a high degree of adaptability, as facilitated by Software Defined Radios (SDRs).
Conversely, the computational aspect requires performance workloads to be adaptable to fluctuating network conditions, necessitating the creation and efficient repeatable testing of innovative computing algorithms.
End-to-end experimentation is essential for obtaining a comprehensive understanding of the system as a whole and for devising innovative approaches to overcome existing bottlenecks.
Additionally, the ability to efficiently replicate these experiments is key to allowing other researchers to easily validate findings.

In this realm of research, we find a diverse landscape of testbeds that have emerged in recent years, each distinguished by its size, applicability, and feature set. 
Notable ones include the
\begin{enumerate*}[itemjoin={{, }}, itemjoin*={{, and }}]
    \item {COSMOS}
    \item {POWDER}
    \item {ARA}
    \item Drexel Grid {SDR}
\end{enumerate*} testbeds.
COSMOS (Cloud enhanced Open Software defined MObile wireless testbed for city-Scale deployment) is a testbed spanning over 1 square mile, equipped with an array of resources, including Software-Defined Radios (SDRs), millimeter-wave (mmWave) equipment, optical fibers, and various computational nodes for core functionality and application processing~\cite{Cosmos1,cosmos2}. 
COSMOS integrates with cloud infrastructure, with a primary focus on advancing cloud computing research. 
This testbed includes a diverse set of rooftop, intermediate, and mobile nodes, all centrally controlled.

POWDER (Platform for Open Wireless Data-driven Experimental Research) serves as an even larger testbed, offering a platform for exploring wireless and mobile networks with a high level of programmability that extends down to the waveform level~\cite{powder}. 
Covering an area of 15 $\mathrm{km}^2$, POWDER comprises fixed, programmable radio nodes constructed using off-the-shelf SDR technology. 
The platform seamlessly integrates with cloud and edge-like compute capabilities, allowing mobile nodes to tap into the compute resources.

The Drexel Grid SDR testbed features SDRs that connect over-the-air (OTA), channel emulator, or alternatively a combination of the two, to enable realistic and reproducible experimentation.
Even though the testbed provides freedom in terms of experiment development, it is intended primarily for SDR-related research, and does not integrate any core, cloud or edge components.

ARA (The Agricultural and ruRAl communities) represents a wireless research platform, spanning a rural expanse with a diameter of 60 km in Iowa~\cite{zhang2022ara}. 
ARA's mission is to investigate wireless platforms and technologies within the real-world context of agricultural settings. 
The testbed incorporates a diverse array of wireless platforms, implemented through SDRs and programmable Commercial Off-The-Shelf (COTS) radios along with automated vehicles and various cameras and other sensors. 
ARA's testbed software builds upon the Chameleon Infrastructure (CHI) software framework from the Chameleon testbed project~\cite{keahey2020lessons}.
Chameleon primarily serves as a bare-metal reconfigurable research testbed designed for edge-to-cloud experimentation, adapting the cloud paradigm for computer science research. 
Since its public unveiling in 2015, Chameleon has accommodated over 6,000 users across more than 800 projects, gaining valuable experience and refining methods for automating and managing research infrastructure sites.
This accumulated knowledge has been encapsulated in the CHI-in-a-Box~\cite{chiinabox} testbed software distribution, which aims to simplify and optimize the deployment and operation of future associated or independent Chameleon sites.

Existing wireless research-focused platforms like COSMOS, POWDER, and Drexel Grid employ their own specialized software solutions, which are not in line with prevailing cloud-native standards. 
These testbeds also utilize virtualization technology based on Virtual Machines (VMs), rather than opting for lighter and more edge-compatible solutions such as containers.
Similar to the approach taken by the ARA testbed, we found that CHI-in-a-Box presents an excellent opportunity to develop experimental platforms at minimal cost, while offering significant flexibility in both cloud and edge-native technologies.
However, when it comes to reproducibility in wireless edge-computing research, existing large-scale testbeds have limitations that we aim to address by introducing our testbed, ExPECA.
By \lq reproducibility\rq, we mean the capacity for researchers to easily replicate experiments and validate results.
Achieving this is particularly challenging for two main reasons: one, edge computing experiments involve a range of heterogeneous components like servers and radios, all interacting within a single experiment; and two, the existence of wireless links introduces variables such as interference and dynamic propagation conditions that must be controlled for consistent results. 
Owing to its unique design choices, ExPECA provides a highly reproducible environment for end-to-end wireless and edge computing experimentation, making it one of the first platforms to effectively address this issue.
In the following section, we delve into these design intricacies and their consequential impacts.

\section{Testbed Design and Architecture}

\begin{figure}[t]
    \centering
    \includegraphics[width=0.99\linewidth]{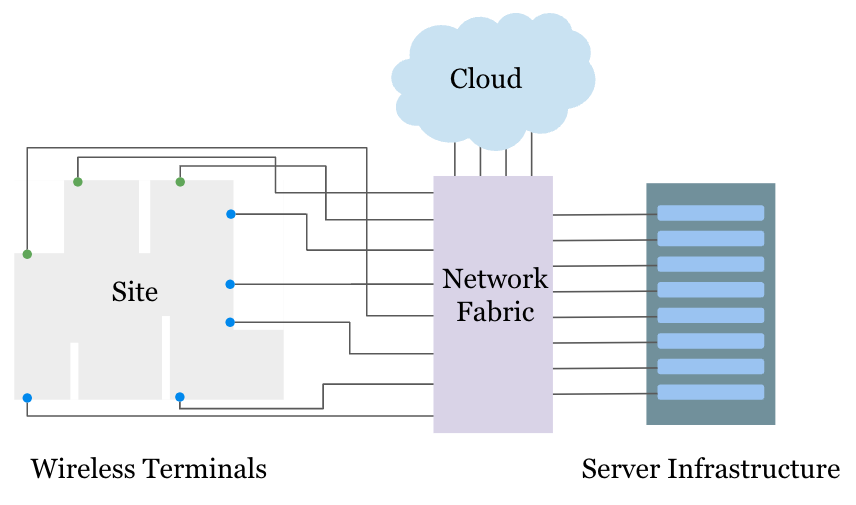}
    \caption{Conceptual architecture of ExPECA for end-to-end edge computing experiments.}\label{fig:concept}
\end{figure}

The ExPECA testbed is designed to address two primary challenges pervasive in edge computing research: 1) end-to-end experimentation, 2) ensuring repeatability and reproducibility.
To enable end-to-end edge computing experiments that encompass both communication and computation aspects, ExPECA is engineered with the following primary goals:
\begin{enumerate}
    \item Provision communication components by utilizing a diverse array of wireless and wired links.
    \item Executing and overseeing computational elements within containerized environments running on the testbed infrastructure.
\end{enumerate}
These objectives are realized through an architectural framework depicted in Figure \ref{fig:concept}.
Central to this architecture is a versatile network fabric that ensures IP connectivity among different experimental components, which may include servers located in the cloud or at the edge, as well as wireless terminals.

Distinctive design choices have been made to realize these objectives, distinguishing ExPECA from contemporary solutions.
We here describe these choices and their impacts.

\subsection{Location and Physical Environment}

\begin{figure}[t]
    \centering
    \includegraphics[width=0.99\linewidth]{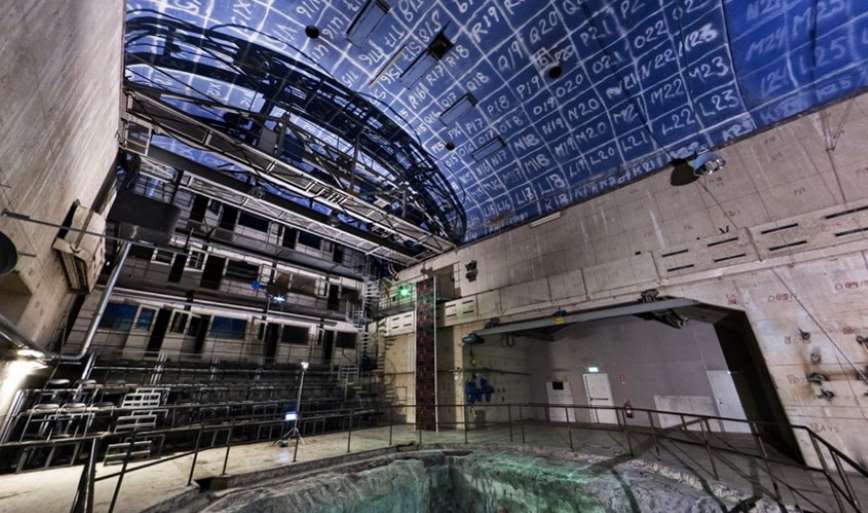}
    \caption{A photograph of the underground site (at KTH R1) of the testbed.}\label{fig:overview}
\end{figure}

The testbed is located in KTH R1 hall, an experimental facility 25 meters below ground. 
This isolated location offers a controlled environment with minimal radio interference, making it ideal for running reproducible wireless experiments.
With dimensions of 12 meters in width, 24 meters in length, and 12 meters in height, the facility is spacious enough to accommodate a variety of remotely controlled devices, such as drones and cars, providing researchers with considerable freedom in designing their experiments.

\subsection{Control and Management Software}

We adopted CHI-in-a-Box as a baseline testbed implementation with extensions and contributions from our team.
CHI-in-a-Box is a packaging of the CHameleon Infrastructure (CHI) software framework which is built primarily on top of the mainstream open-source OpenStack platform. 
CHI has mature support for edge-to-cloud provisioning of compute, storage, and network fabrics, and provides end-to-end experiment control and user services.
Specifically, we leverage the CHI@Edge flavor of the framework, a novel version that supports running containerized workloads on compute resources.
In the following, we describe the essential elements of CHI.
Users have the flexibility to allocate resources either on-the-fly or through advance reservations, facilitated by customized versions of OpenStack's Blazar and Doni services. 
These services have been tailored by the ExPECA team to include unique features such as radio and Kubernetes node reservations. 
The scope of allocatable resources encompasses worker nodes, radios, networks, and IP addresses. 
Once resources are secured, the integrated Kubernetes orchestration system enables the deployment of containerized workloads via OpenStack Zun. 
Users have the option to utilize pre-configured Docker images supplied by the ExPECA team (as in the case of SDR RAN) or to upload their own container images. 
Notably, the ExPECA team has extensively modified Zun to support functionalities like network interface and block storage attachments to K8S containers, features not originally supported in CHI@Edge.
Configuration of complex experiment topologies such as distributed networking experiments is supported through CHI's programmable interface and our adopted flavor of the \texttt{python-chi} Python library.
In addition to these, OpenStack Neutron plays a pivotal role in managing the network architecture. Neutron provides a robust set of networking capabilities, including virtual networking, that allows for the creation of isolated networks, subnets, and routers.
We connect all testbed components, including radios, to managed Layer 2 switches controlled by Neutron.

\subsection{Radio Nodes}
\begin{figure}[t]
  \begin{subfigure}{0.3\linewidth}
    \centering
    \includegraphics[width=1\linewidth]{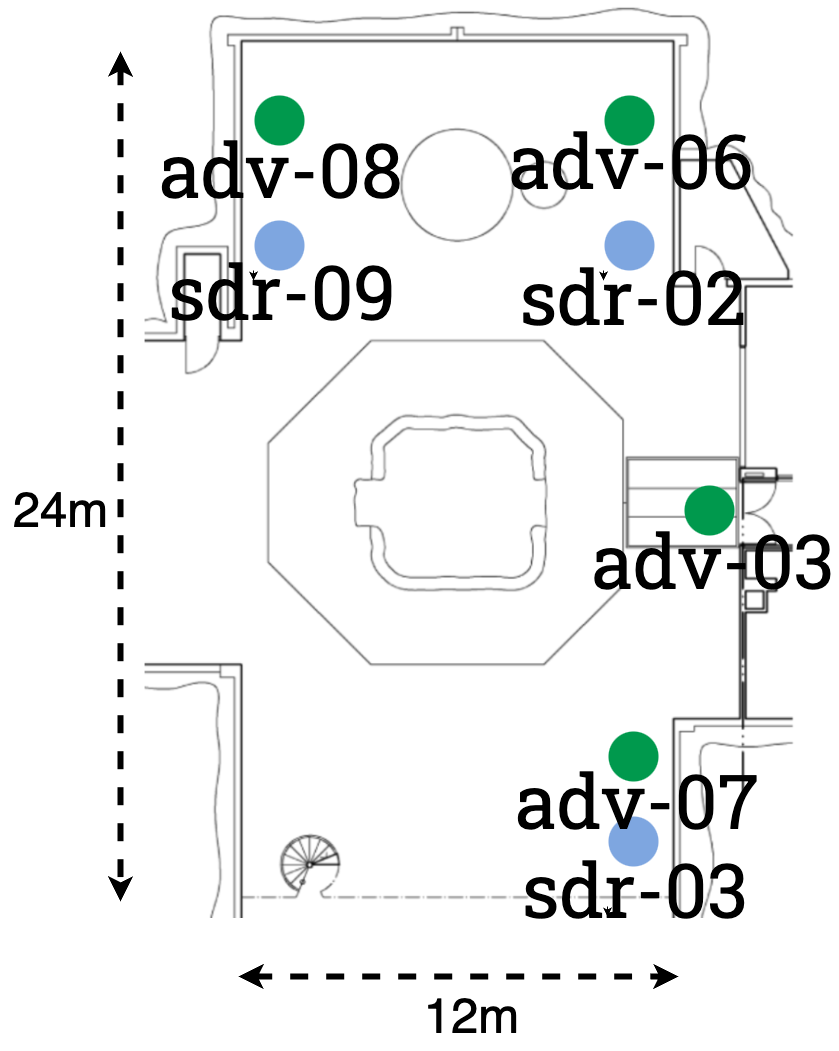}
    \caption{2m high}
  \end{subfigure}
  \begin{subfigure}{0.3\linewidth}
    \centering
    \includegraphics[width=1\linewidth]{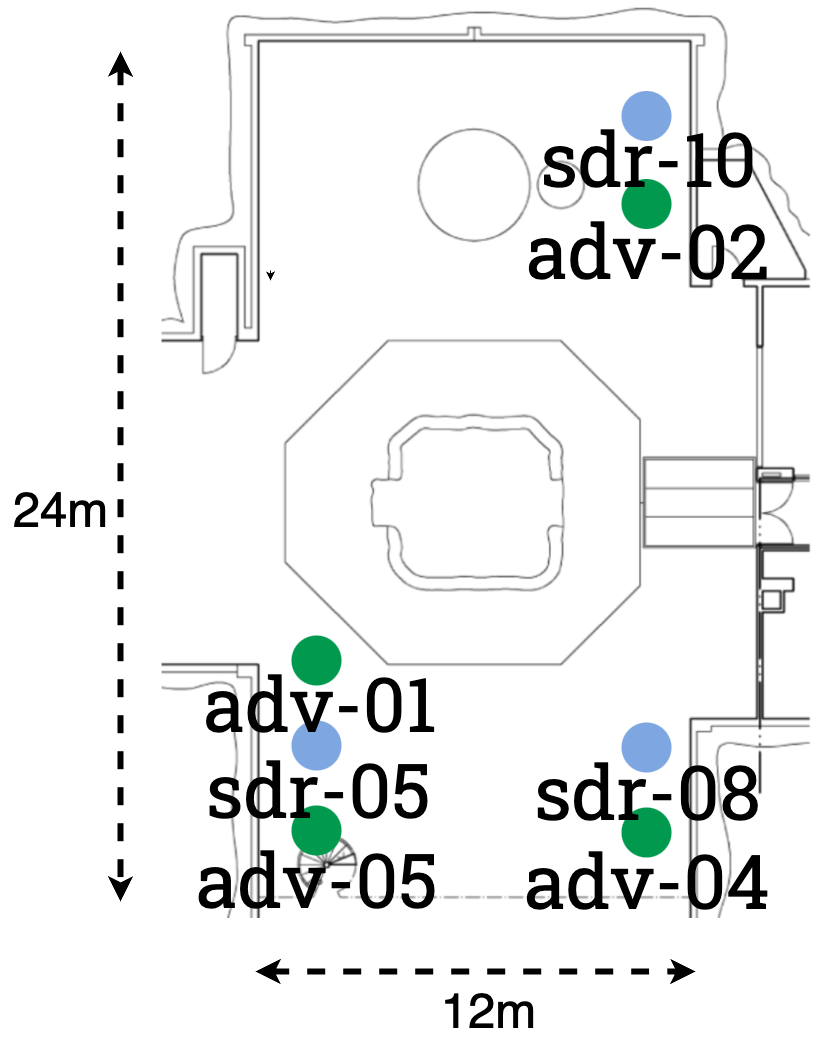}
    \caption{5m high}
  \end{subfigure}
  \begin{subfigure}{0.3\linewidth}
    \centering
    \includegraphics[width=1\linewidth]{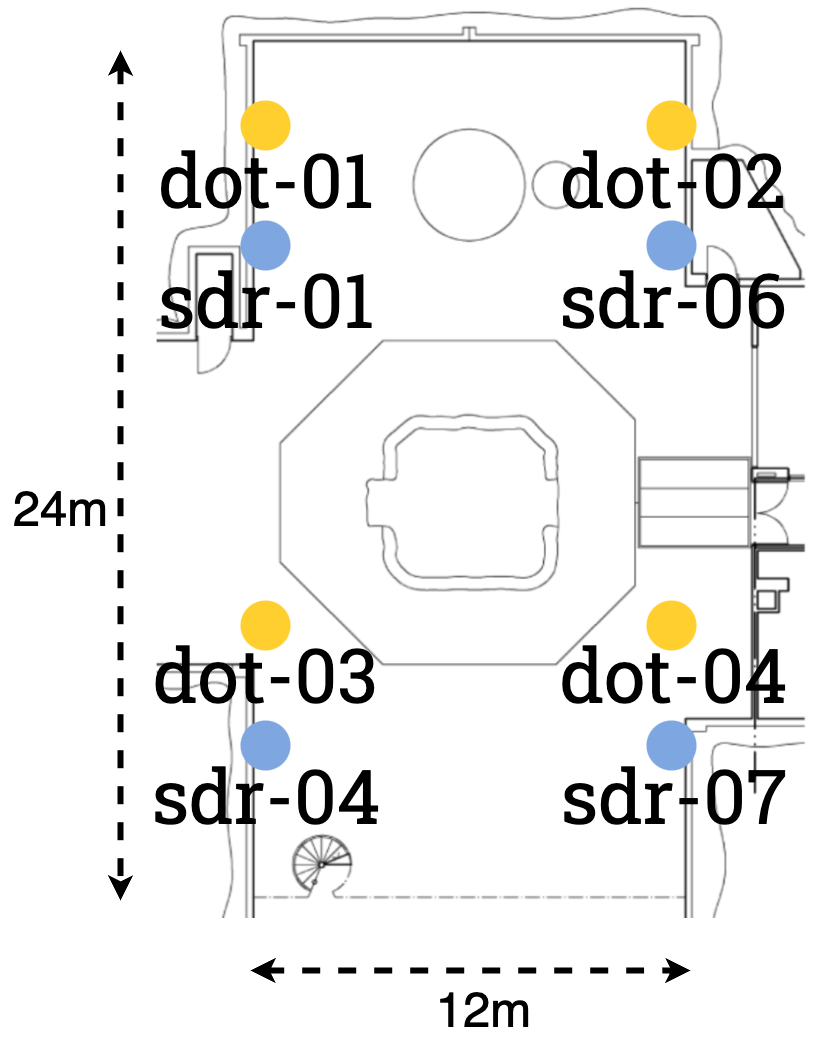}
    \caption{10m high}
  \end{subfigure}
  \caption{Low, mid, and high-level radio nodes locations at R1 on three different levels as of September 2023.}
  \label{fig:radiosmap}
\end{figure}

One of the standout features of ExPECA is the inclusion of wireless terminals as reservable resources.
These radios are mounted at diverse locations in R1, at three different levels, as shown in Figure \ref{fig:radiosmap}. 
They are categorized into two types:
\begin{enumerate}
\item SDR Nodes: These can serve as the air interface for any wireless protocol implemented for USRP E320 SDRs, including OpenAirInterface (OAI) 5G, OAI LTE, SRS LTE, Mangocomm IEEE802.11b/g, or any GNU radio-based wireless protocol.
\item COTS Radios: Includes an Ericsson private 5G system with radio dots and Advantech routers serving as 4G/5G COTS UE nodes.
\end{enumerate}
We chose all radios, specifically SDRs, to be equipped with IP-networked interfaces.
This allowed us to enroll them as network segments in Openstack Blazar and Doni with minimal extension efforts.
The users can easily integrate them into any workload container or SDR host container, through the testbed software.
For example, researchers can seamlessly transition from a high-quality 5G link between locations sdr-01 and sdr-06 to a poor-quality channel between sdr-09 and sdr-07, enabling them to empirically assess the impact on application performance.
Alternatively, they can switch to a WiFi link between sdr-01 and sdr-06 for further experimentation.

\subsection{Compute Nodes}

ExPECA offers compute resources by allowing users run containerized applications with the most recent version of Openstack Zun that uses K8S (Kubernetes) as the orchestration layer.
K8S is designed to handle a large number of containers across a variety of environments, making it well-suited for complex, large-scale deployments. 
Also, it offers a rich set of APIs and a pluggable architecture, allowing for greater customization and extensibility.
This was particularly useful for our team when developing new desired features in Zun.
These new features include 1) network attachment, 2) block storage attachment, 3) resource management i.e. specifying the number of cores and memory size for the container.
Users are able to run their containerized applications on ExPECA's Intel x86 servers, enrolled in the K8S cluster, which we refer to as 'workers', not only to deploy and orchestrate application workloads, but also to deploy SDR communication stacks or apply radio node configurations.
With the advent of Open Radio Access Network (O-RAN) and SDRs, all components of the wireless network stack can run on general-purpose processors.
They can thus be containerized and distributed across multiple hosts as long as the containers are connected to the SDRs.
We provide reference container images for SDR 5G, SDR LTE, and WiFi provisioning.
Lastly, it is important to note that all worker nodes achieve sub-$\mu$s synchronization by exchanging PTP messages with a Grandmaster (GM) clock that itself uses GPS as a reference. Tight time synchronization is essential for users who require the same timing reference to timestamp packets across various locations within the testbed.

\section{Experimental Testbed Validation}

This section aims to substantiate two key assertions about ExPECA: its efficiency in facilitating end-to-end experiments and its ability to ensure experimental reproducibility\footnote{Experiments available as Python Jupyter notebooks at https://github.com/KTH-EXPECA/examples}.

\begin{figure}
\centering
\begin{subfigure}{0.91\linewidth}
    \begin{center}
        \includegraphics[width=1\linewidth]{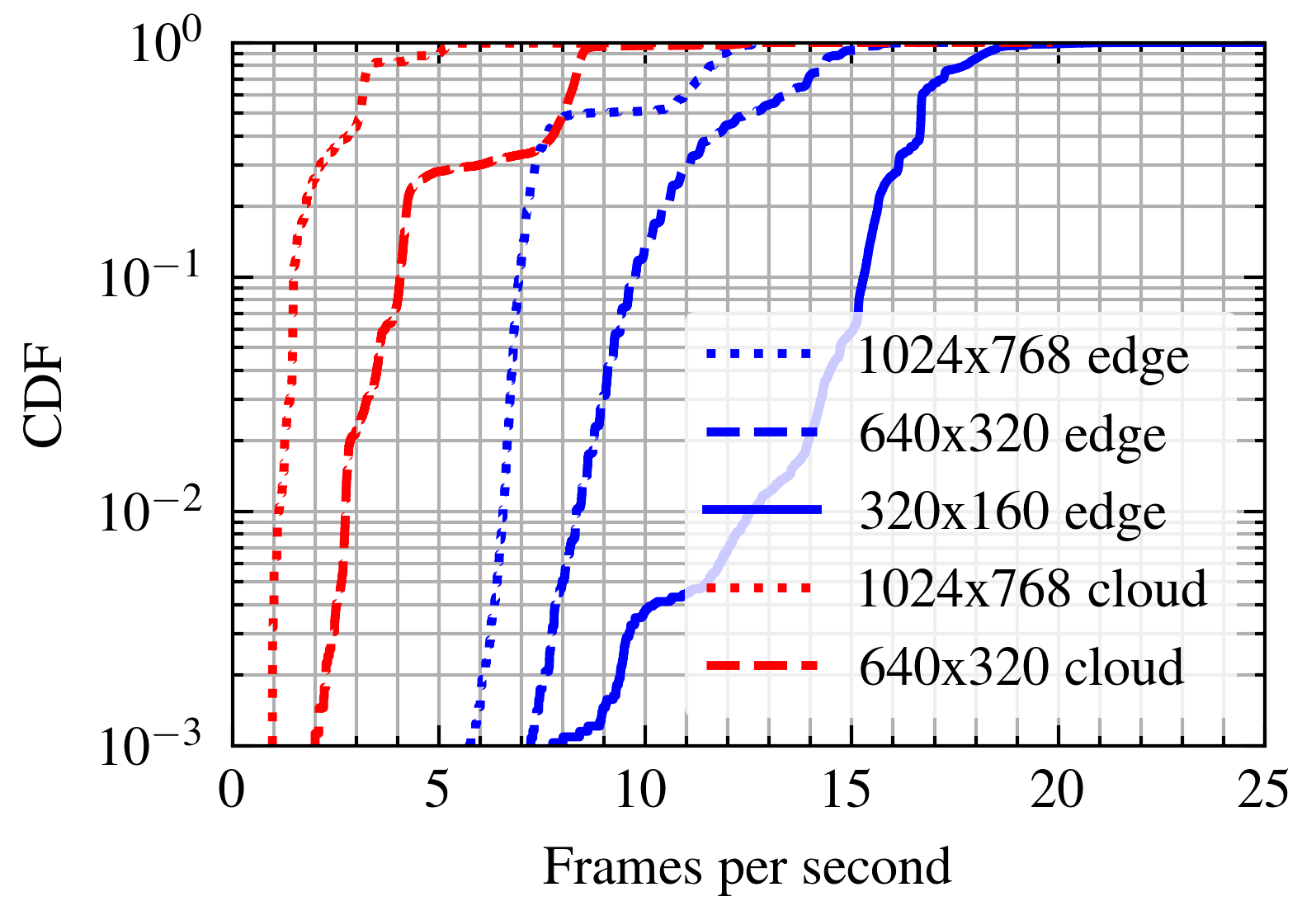}
    \end{center}
    \caption{CDF of FTT.}
    \label{fig:openrtist_A}
\end{subfigure}
~
\begin{subfigure}{0.91\linewidth}
  \begin{center}
        \includegraphics[width=0.91\linewidth]{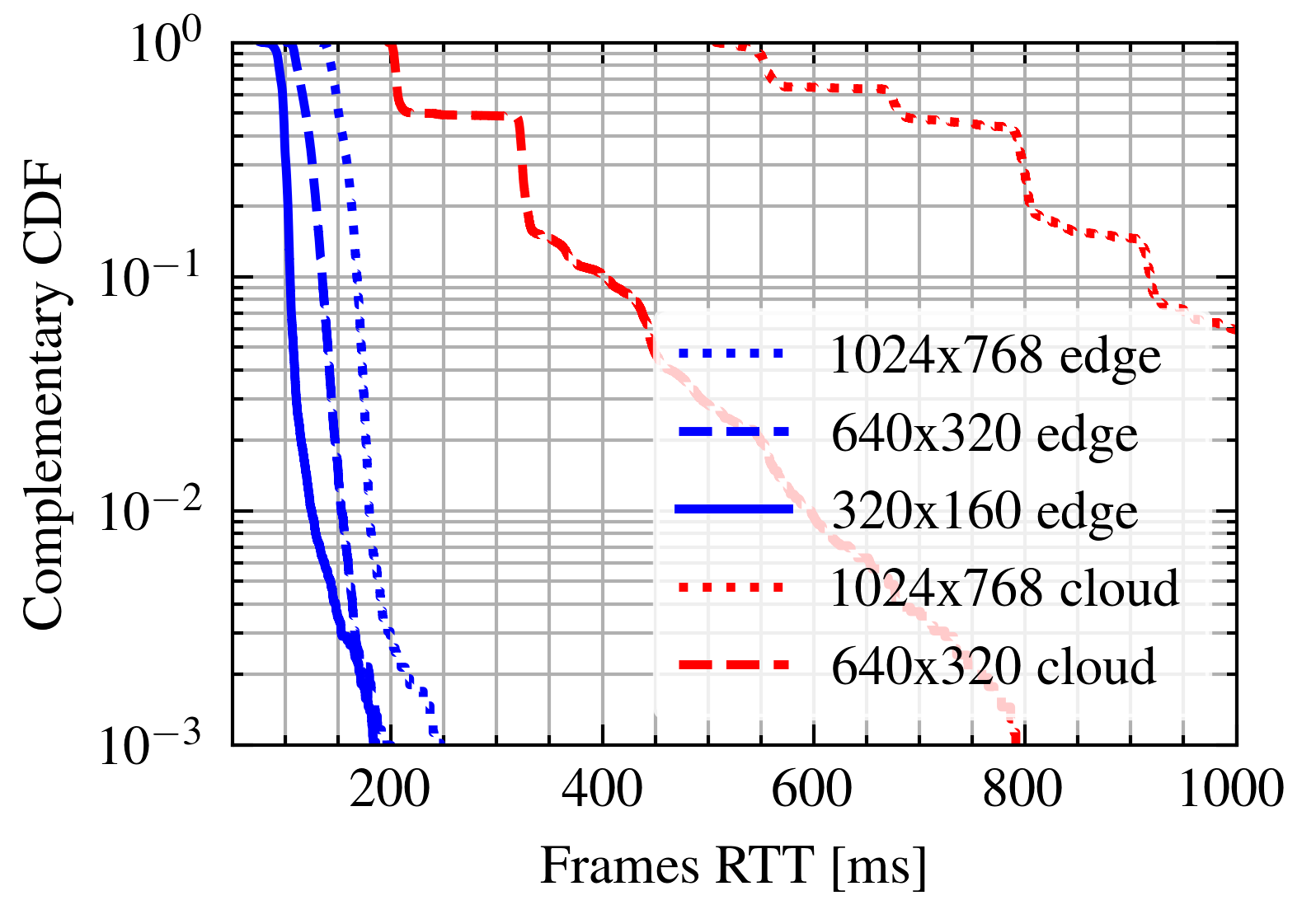}
    \end{center}
    \caption{CCDF of RTT.}
    \label{fig:openrtist_B}
\end{subfigure}
\caption{Performance analysis of OpenRTiST on COTS 5G system and adv-03 node using the ExPECA testbed.} \label{fig:openrtist}
\end{figure}

To validate the first claim, we employ OpenRTiST~\cite{openrtist}, an open-source application that is computationally demanding, bandwidth-intensive, and latency-sensitive. 
OpenRTiST uses machine learning filtering to overlay various artistic styles onto video frames, creating an augmented reality experience.
In this experiment, the OpenRTiST client connects to the server either at the edge, specifically the testbed servers, or in the cloud via a server located at Carnegie Mellon University (CMU), in Pittsburgh, Pennsylvania, through a COTS 5G system and the adv-03 node.
In Figure~\ref{fig:openrtist}, we depict the corresponding critical performance indicators such as frames per second (FPS) and round-trip time (RTT) measured using the testbed. In Figure~\ref{fig:openrtist_A}, we plot the Cumulative Distribution Function (CDF) of FPS, and in Figure~\ref{fig:openrtist_A} we plot the Complementary CDF (CCDF) of RTT.
As we move from edge to cloud, RTT significantly increases which results in lower refresh rates and less responsive video stream. 
The performance is benchmarked against different video frame sizes to analyze network and compute capacity.

\begin{figure}[t]
    \centering
    \includegraphics[width=0.91\linewidth]{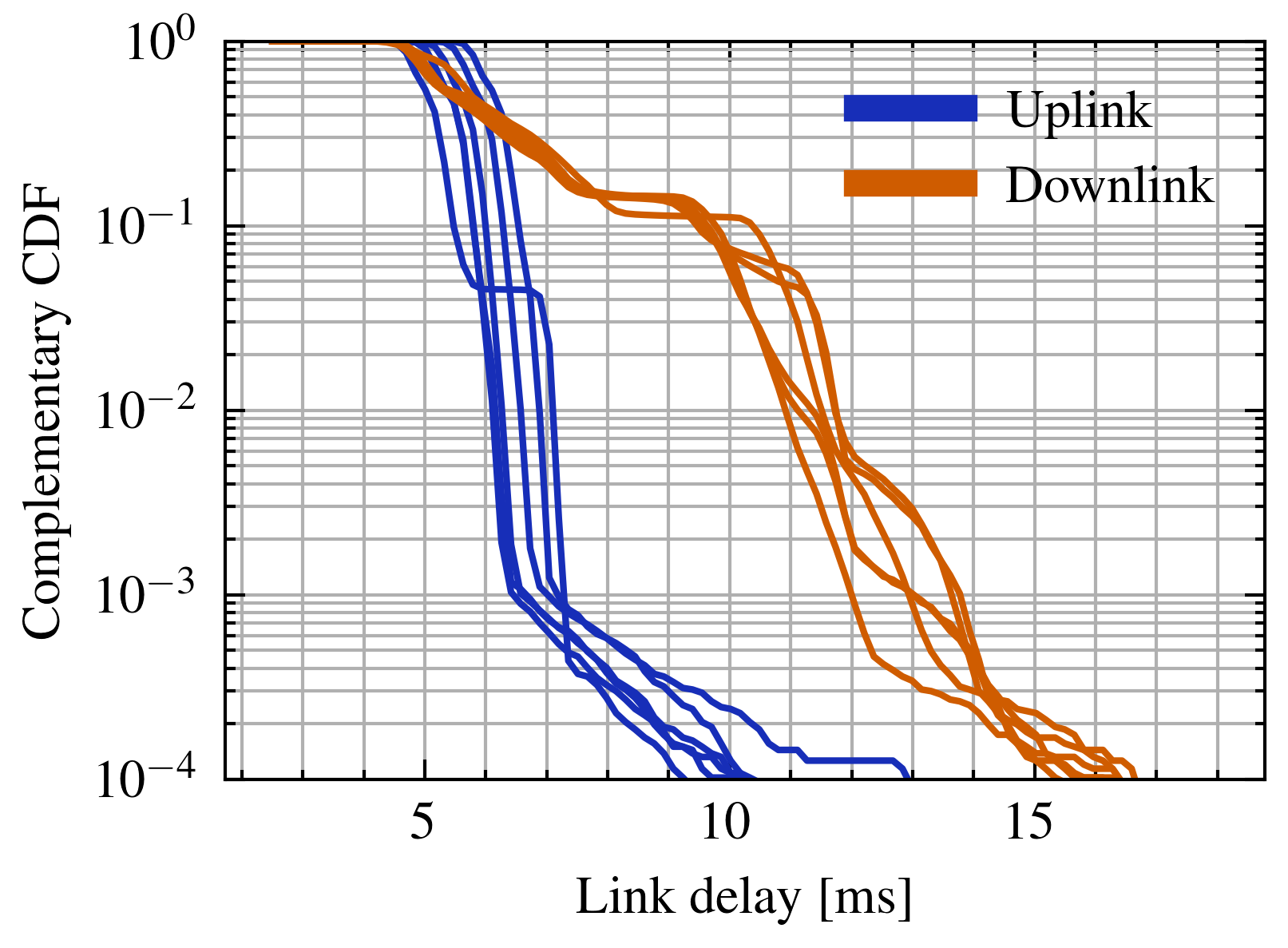}
    \caption{Twelve rounds of independent end-to-end latency measurements over the 5G COTS system between dot-02 and adv-03 node, each based on 160k samples.}\label{fig:ep5grep}
\end{figure}

For the second claim, we concentrate on wireless network latency, a metric inherently challenging to model due to its stochastic nature at the packet level. 
While latency can be influenced by factors like transmit power and channel interference, it also depends on wireless protocol settings such as modulation schemes, bandwidth, and antenna configurations. 
ExPECA's isolated location minimizes variability in channel quality, mitigating potential interference. 
In Figure~\ref{fig:ep5grep}, we present latency measurements across different radio links over several days, demonstrating the reproducibility level in the case of COTS 5G wireless links.
The close similarity between these CCDFs confirms the reproducibility of wireless latency measurements achieved through ExPECA.

\section{Supported Experimentation}

\begin{figure}[t]
    \centering
    \includegraphics[width=0.9\linewidth]{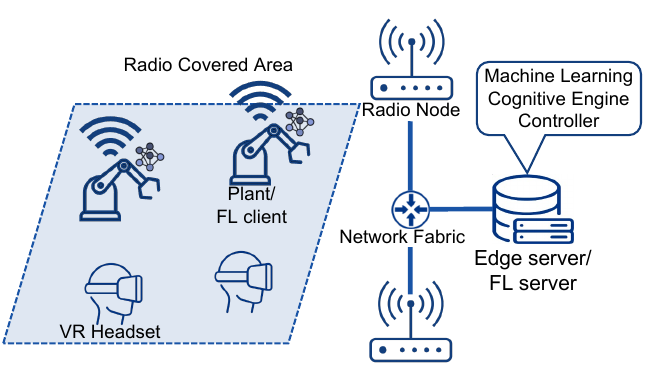}
    \caption{Illustration: Testbed experiment scenarios.}\label{fig:examples}
\end{figure}

This section delves into the features of the testbed that users can customize to meet their unique experimental goals. 
These features can be broadly categorized into two types: (1) configurations related to workload and (2) configurations related to the network.

In terms of workload-related configurations, ExPECA offers researchers a range of options to tailor to their experiments. 
Users can select from various geographical locations for computational workloads using the cloud interface and by reserving public IPs. 
It is possible to integrate GPUs for tasks requiring accelerated processing, or choose a CPU with a certain number of cores, and allocate a specific amount of RAM to benchmark workload performance.
The platform supports experiments involving both static (fixed-location) and mobile (varying-location, e.g., drone with a wireless connectivity dongle) wireless end nodes.

For network-related configurations, ExPECA provides a diverse array of network types, such as SDR 5G, SDR LTE, SDR WiFi, and COTS 5G, among others. 
Researchers can also customize network topologies and control interference levels, thanks to the testbed's isolated location.
The platform supports a wide range of wireless protocols compatible with USRP e320 and allows users to configure channel conditions. Additionally, the use of Software Defined Radios (SDRs) offers further flexibility, including control over transmit power and the ability to implement a variety of wireless protocols.

ExPECA offers the opportunity to explore various experimentation scenarios involving edge computing, wireless communication, and machine learning research.
Here we explain various scenarios, each shedding light on some of the interesting aspects of wireless communication and/or edge computing offering unique insights into the dynamic landscape of edge computing. In Figure~\ref{fig:examples}, we provide an illustration that helps a reader to visualize these scenarios.
Below, we provide a few of such use cases in various research areas.
While some of these aim at direct measurements for data-driven modeling and prediction, others target initial observations for forthcoming theoretical work or the verification and validation of existing theoretical models. 

\subsection{Optimization of Closed-Loop Applications}
With the booming interest and applications surrounding cyber-physical systems, it is imperative to study Networked Control Systems (NCSs)~\cite{8766208}. 
NCSs form crucial control systems where the control loops are closed through a communication network. 
Due to the safety-critical nature of the applications these NCSs encompass, such as robots, UAVs etc., there is a need to benchmark and understand the bottlenecks/boundaries associated with these applications~\cite{olguinmunoz2023phdthesis}. 
Namely, the latency and reliability boundaries of robotic control are crucial to understand before they can be deployed in real-world scenarios, where their potential failure might prove expensive~\cite{cleave}. 

Another such feedback system where latency is of importance are human-in-the-loop applications such as augmented reality,
wearable cognitive assistants (WCA)~\cite{TowardsWCA,olguinmunoz2023phdthesis}, and ambient safety. 
In the context of criticality, we have systems of automated fault detection, in which case the acoustic~\cite{acoustic} or motion amplified visual~\cite{oh2018learning,liu2005motion} data is processed for vibration analysis to potentially trigger some maintenance, safety or emergency procedures~\cite{acoustic}.
The ExPECA testbed provides developers with the opportunities to experiment with workloads corresponding to these types of applications in a containerized manner with various real network architectures.
These experiments could either be for data collection or for verification of existing theoretical models.

\subsection{Time-sensitive Wireless Networking}
\begin{figure}[t]
    \centering
    \includegraphics[width=0.99\linewidth]{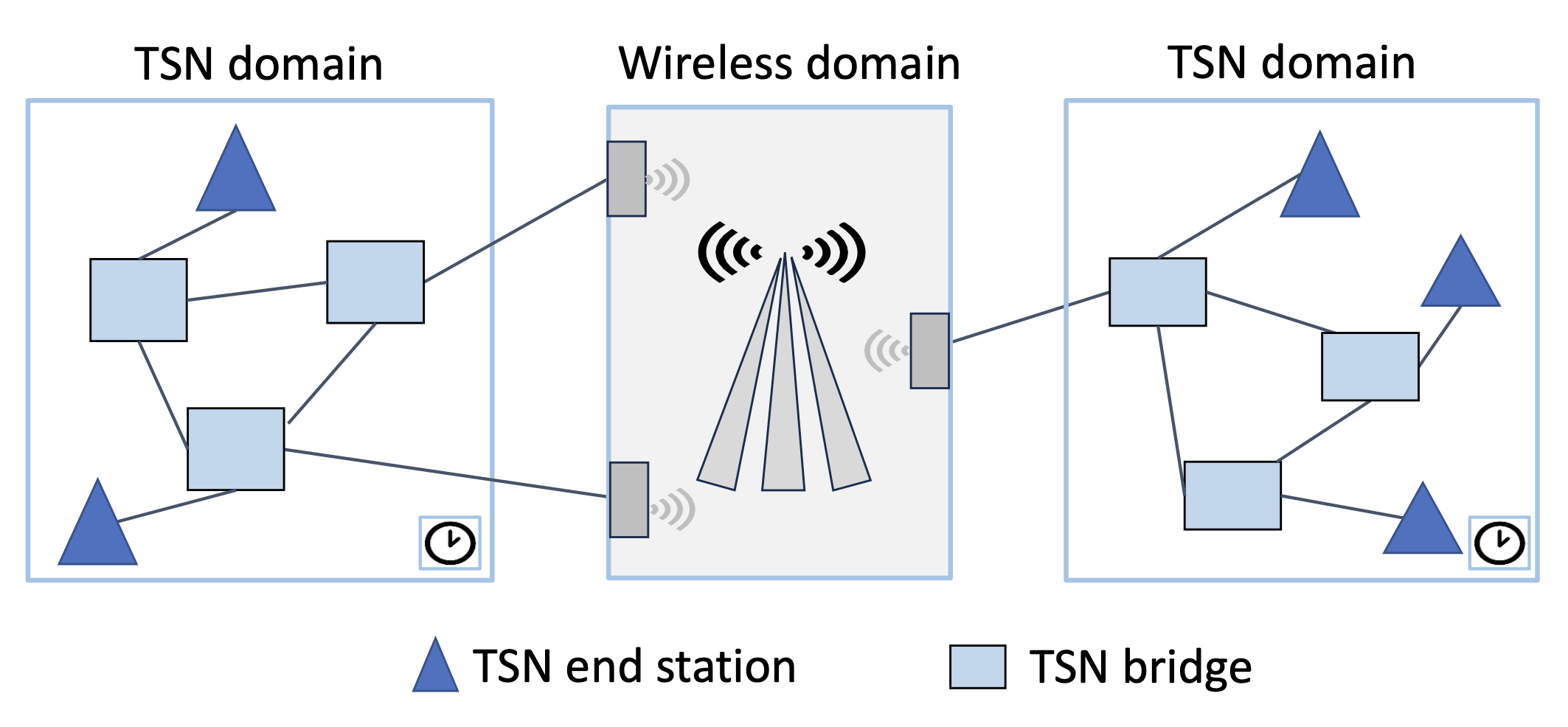}
    \caption{An illustration for the integration of a wireless domain with TSN domains.}\label{fig:wireless-tsn-int}
\end{figure}

Supporting the above-mentioned CPS and HITL applications requires the network to offer a certain level of \textit{end-to-end} network latency, often combined with extremely high reliability, in order to operate correctly and safely. 
This has spurred various efforts to integrate wireless networking technologies (e.g., 5G URLLC) with wired technologies (e.g., TSN) as shown in Figure \ref{fig:wireless-tsn-int}. 
However, an efficient wired-wireless integration realization to support time-critical applications necessitates accurate system characterization. 
In particular, latency characterization of wireless segments is crucial as they are subject to various stochastic and temporal influences, unlike wired links where latencies are immune to external influences. 
Furthermore, accurate predictions of wireless latency are also envisioned to be leveraged in 6G networks to support deterministic communications \cite{sharma2023towards}.

Lately, several data-driven methods for latency prediction have been proposed \cite{mostafavi2023data,skocaj2023data}. 
Training and validating these approaches require a vast amount of datasets comprising latency measurements alongside network parameters (e.g., SINR and MCS). 
Precise time-synchronization of all nodes of the testbed ensures accurate latency measurements can be collected between any two nodes of the testbed. 
These experiments can employ diverse wireless communication technologies, including COTS 5G, software-defined 5G and IEEE 802.11g Wi-Fi integrated in ExPECA. 
Furthermore, the containerized workflow within ExPECA enables users to perform automated, reproducible and repetitive latency measurement experiments.

\subsection{Hierarchical Learning and Inference}
Another set of experiments researchers can use the ExPECA testbed for is the verification of theoretical results corresponding to Hierarchical Learning (HI)~\cite{hi1,hi2}. 
In these works, authors develop the idea of Hierarchical Inference, where a compute constrained edge device chooses to perform local inference or classification of an image sample, or offloading via a wireless channel to get help of a high power edge server for classification.
A small-sized ML model with limited accuracy is assumed to have been deployed at the edge device and offloading to the server incurs cost. 
In essence, these policies aim to make intelligent decisions regarding task offloading, prioritizing complex tasks or tasks where accuracy is paramount. 
Conversely, they process simpler tasks or tasks where reduced latency is of utmost importance locally. 
It is to be noted that in this specific work on Hierarchical Inference this decision is made in an online manner. 
Using the ExPECA testbed, researchers can plan to implement the various algorithms proposed and verify if the real-world results match simulations and models.

\subsection{Federated Learning Validation}
Federated learning (FL) is emerging as one of the key approaches for training machine learning models~\cite{GoogleFL}. 
FL aims to improve over centralized learning in terms of privacy, total training time, computation overhead, etc. 
However, FL can cause large communication overhead because the machine learning model should be repeatedly exchanged between the central server and the participating nodes during the training process. 
In order to analyze and optimize communication overhead, many works have proposed models for FL systems operating in wireless environments ~\cite{fl1,fl2}.
Validating these theoretical models in real systems is required to reveal otherwise unavailable insights.
The ExPECA testbed supports experimentation involving both centralized learning (CL) and federated learning to evaluate them in terms of communication/computation overhead and accuracy with various wireless conditions.
For example in case of communication overhead, CL and FL will consume communication resources when exchanging data (i.e., using uplink to send input data in CL and uplink/downlink to send machine learning model in FL).
The ExPECA testbed is able to measure the communication resource usage, and training time (communication/computation time), and capture wireless parameters at that time. 
Thanks to improved reproducibility and repeatability, the ExPECA testbed supplies stable validation for machine learning architectures.

\section{Acknowledgements}
This research has been partially funded by (1) the VINNOVA Competence Center for Trustworthy Edge Computing Systems and Applications (TECoSA) at KTH Royal Institute of Technology; and (2) the Swedish Foundation for Strategic Research (SSF), through
grant number ITM17–0246.

\bibliographystyle{ieeetr}
\bibliography{refs.bib}

\end{document}